\documentclass[article]{llncs}

\usepackage[T1]{fontenc}

\usepackage{fancyhdr}
\usepackage{plain}
\usepackage[pdftex]{graphicx}
\usepackage{caption}
\usepackage{amssymb}
\usepackage{amsmath}
\usepackage{multirow}
\usepackage{tablefootnote}
\usepackage{float}
\usepackage[obeyspaces,spaces]{url}
\usepackage{cite}
\usepackage{todonotes}
\usepackage{algpseudocode}
\usepackage{paralist}
\usepackage{booktabs}
\usepackage{subfig,array,ragged2e}
\usepackage{threeparttable}
\usepackage{listings}
\usepackage{longtable}
\usepackage{lscape}
\usepackage{xspace}
\usepackage{enumitem}
\usepackage{ntheorem}

\usepackage{hyperref}
\usepackage{backref}

\newcommand{\ie}{i.e.}
\newcommand{\eg}{e.g.}
\def\TestREx{\textsc{TestREx}\xspace}
\long\def\longcaption#1#2{\centering\begin{minipage}{#1}\small\emph{#2}\end{minipage}}

\begin{document}
\sloppy

\title{\TestREx: a Framework for Repeatable Exploits}

\author{Stanislav Dashevskyi\inst{1},
		Daniel Ricardo dos Santos\inst{1,2},
		Fabio Massacci \inst{1}, and
	    Antonino Sabetta \inst{3}
}
\institute{
	University of Trento, Italy  	\email{stanislav.dashevskyi@unitn.it, fabio.massacci@unitn.it}\and
	Fondazione Bruno Kessler, Italy \email{dossantos@fbk.eu} \and
	SAP Labs, France 				\email{antonino.sabetta@sap.com}
}

{\def\addcontentsline#1#2#3{}\maketitle}

{\large
This draft was submitted to arXiv.org\footnote{\url{https://arxiv.org/}}.  The
final publication is available at Springer via
\url{http://dx.doi.org/10.1007/s10009-017-0474-1}.
}

\title{\TestREx: a Framework for Repeatable Exploits~\thanks{This
paper is an extension of our previous work~\cite{dashevskyi2014testrex}}}

\author{Stanislav Dashevskyi\inst{1},
		Daniel Ricardo dos Santos\inst{1,2},
		Fabio Massacci \inst{1}, and
	    Antonino Sabetta \inst{3}
}
\institute{
	University of Trento, Italy  	\email{stanislav.dashevskyi@unitn.it, fabio.massacci@unitn.it}\and
	Fondazione Bruno Kessler, Italy \email{dossantos@fbk.eu} \and
	SAP Labs, France 				\email{antonino.sabetta@sap.com}
}

\setcounter{tocdepth}{3}
\tableofcontents

{\def\addcontentsline#1#2#3{}\maketitle}

\begin{abstract}
Web applications are the target of many well known exploits and also a
fertile ground for the discovery of security vulnerabilities. Yet, the
success of an exploit depends \emph{both} on the vulnerability in the
application source code \emph{and} the environment in which the application
is deployed and run. As execution environments are complex (application
servers, databases and other supporting applications), we need to have a
reliable framework to test whether known exploits can be reproduced in
different settings, better understand their effects, and facilitate the
discovery of new vulnerabilities.  In this paper, we present \TestREx\ -- a
framework that allows for highly automated, easily repeatable exploit testing
in a variety of contexts, so that a security tester may quickly and
efficiently perform large-scale experiments with vulnerability exploits. It
supports packing and running applications with their environments, injecting
exploits, monitoring their success, and generating security reports. We also
provide a corpus of example applications, taken from related works or
implemented by us.

\end{abstract}


\section{Introduction}
\label{sec:intro}

Web applications are nowadays one of the preferred ways of providing services
to users and customers. Modern application platforms provide a great deal of
flexibility, including portability of applications between different types of
execution environments, e.g., in order to meet specific cost, performance, and
technical needs.  However, they are known to suffer from potentially
devastating vulnerabilities, such as flaws in the application design or code,
which allow attackers to compromise data and functionality (see,
e.g.,~\cite{stuttard2007,zalewski2011}). Vulnerable web applications are
a major target for hackers and cyber attackers~\cite{scholte2011}, while
vulnerabilities are hard to identify by traditional black-box approaches for
security testing~\cite{dilucca2006,curphey2006,tripp2014}.

A key difficulty is that web applications are deployed and run in many
different execution environments, consisting of operating systems, web servers,
database engines, and other sorts of supporting applications in the backend, as
well as different configurations in the frontend~\cite{dilucca2006}. Two
illustrative examples are SQL injection exploits (success depends on the
capabilities of the underlying database and the authorizations of the user who
runs it~\cite[Chapter 9]{stuttard2007}), and Cross-site Scripting (XSS)
exploits (success depends on the browser being used and its rules for executing
or blocking JavaScript code~\cite[Chapter 14]{zalewski2011}). These different
environments may transform failed attempts into successful exploits and vice
versa.

Industrial approaches to black-box application security testing (e.g., IBM
AppScan\footnote{\url{http://www.ibm.com/software/products/en/appscan}}) or
academic ones (e.g., Secubat~\cite{kals2006} and BugBox~\cite{nilson2013})
require security researchers to write down a number of specific exploits that
can demonstrate the (un)desired behavior. Information about the configuration
is an intrinsic part of the vulnerability description. Since the operating
system and supporting applications in the environment can also have different
versions, this easily escalates to a huge number of combinations which can be
hard to manually deploy and test.

We need a way to automatically switch configurations and re-test exploits to
check whether they work with a different configuration. Such data
should also be automatically collected, so that a researcher can see how
different exploits work once the configuration changes. Such automatic process
of ``set-up configuration, run exploit, measure result'' was proposed by Allodi
et al.~\cite{allodi2013malwarelab} for testing exploit kits, but it is not
available for testing web applications.

Our proposed solution,
\TestREx\footnote{\url{http://securitylab.disi.unitn.it/doku.php?id=testrex}},
combines packing applications and execution environments that can be easily and
rapidly deployed, scripted exploits that can be automatically injected, useful
reporting and an isolation between running instances of applications to provide
a real ``playground'' and an experimental setup where security testers and
researchers can perform their tests and experiments, and get reports at various
levels of detail.

We also provide a corpus of vulnerable web applications to illustrate the usage
of \TestREx over a variety of web programming languages. The exploit corpus is
summarized in Table~\ref{tab:summary:exploits}. Some of the exploits are taken
from existing sources (e.g., BugBox~\cite{nilson2013} and
WebGoat~\cite{webgoat}), while others are developed by us. For the latter
category, we focused on server-side JavaScript, because of its growing
popularity in both open source and industrial usage (e.g.,
Node.js\footnote{\url{http://nodejs.org/}} and SAP
HANA\footnote{\url{https://help.sap.com/hana}}) and, to the best of our
knowledge, the lack of vulnerability benchmarks.

\begin{table}[!ht]
\caption{Available exploits in \TestREx corpus}
\longcaption{\columnwidth}{\\}
\label{tab:summary:exploits}		
\centering
\begin{tabular}{| c | c | c |}
\hline
	\textbf{Language} & \textbf{Exploits} & \textbf{Source} \\
\hline
	PHP & 83 & BugBox~\cite{nilson2013} \\
\hline
	Java & 10 & WebGoat~\cite{webgoat} \\
\hline
	Server-side JavaScript & 7 & TestREx \\
\hline
\end{tabular}
\end{table}

The rest of this paper is organized as follows. Section~\ref{sec:related:work}
describes and compares related work in the field of security experimentation;
Section~\ref{sec:overview} presents an overview of \TestREx;
Section~\ref{sec:implementation} discusses the implementation of the framework;
Section~\ref{sec:evaluation} describes our evaluation of \TestREx with testing
various exploits, as well as using it as an educational tool;
Section~\ref{sec:usage:model} lists potential uses of \TestREx, focusing on an
industrial context; finally, Section~\ref{sec:conclusion} concludes the paper
with the main lessons learned and a brief description of future work ideas. In
the Appendix we provide a detailed guide on how to contribute to \TestREx.

\section{Related Work}
\label{sec:related:work}

Security testing verifies and validates software requirements related to
security properties such as confidentiality, integrity, and availability.
Felderer et al.~\cite{felderer2016} argue that support for security testing is
essential to increase its effectiveness and efficiency in practice.

Such support requires the development of experimental frameworks where
developers can actually test and experiment with security bugs as advocated
initially in~\cite{eide2010toward,maxion_killouhry2011,carroll2012}. Yet, this
is far from trivial and only few papers in mainstream security conferences use
experiments as their validation measure (e.g., at IEEE S\&P 2015 only 3 papers
out of 55 used experiments~\cite{carver2016}).

Indeed, a number of issues should be tackled in order to correctly provide
security experimentation setups. These issues include isolation of the
experimental
environment~\cite{benzel2011,allodi2013malwarelab,calvet2010isolated,nilson2013},
repeatability of individual
experiments~\cite{eide2010toward,allodi2013malwarelab}, collection of
experimental results, and justification of collected
data~\cite{maxion_killouhry2011}.

The use of a structured testbed can help in achieving greater control over the
execution environment, isolation among experiments, and reproducibility. Most
proposals for security research testbeds focus on the network level (e.g.,
DETER~\cite{benzel2011}, ViSe~\cite{arnes2006}, and vGrounds~\cite{jiang2006}).
A comparison of network-based experimental security testbeds can be found
in~\cite{stoner2015}. On the application level there are significantly less
experimental frameworks.

By using the taxonomy of Felderer et al.~\cite{felderer2016}, the security 
testing process must be present in every phase of the lifecycle: model-based 
security testing between analysis and design; code-based testing and static 
analysis during development; penetration testing and dynamic analysis during 
deployment; and regression testing during maintenance. Application-based 
security experimental frameworks such as \TestREx, BugBox, or WebGoat support 
security testers at the later stages of the lifecycle, namely deployment and
maintenance.

Among the application level-frameworks, the BugBox framework~\cite{nilson2013}
provides the infrastructure for deploying vulnerable PHP-MySQL web 
applications, creating exploits and running these exploits against applications 
in an isolated and easily customizable environment. As in BugBox, we use the 
concepts of execution isolation and environment flexibility. However, we needed 
to have more variety in software configurations and process those 
configurations  automatically. We have broadened the configurations scope by 
implementing software images for different kinds of web applications, and 
automatically deploy them.

The idea of automatically loading a series of clean configurations every time
before an exploit is launched was also proposed by Allodi et al. in their
MalwareLab~\cite{allodi2013malwarelab}. They load snapshots of virtual machines
that contain clean software environment and then ``spoil'' the environment by
running exploit kits. This eliminates the undesired cross-influence between
separate experiments and enforces repeatability. So we have incorporated it
into \TestREx. For certain scenarios, cross-influence might be a desired
behavior, therefore \TestREx makes it possible to run an experiment suite in
which the experimenter can choose to start from a clean environment for each
individual exploit/configuration pair or to reuse the same environment for a
group of related exploits.

Maxion and Killourhy~\cite{maxion_killouhry2011} have shown the importance of
comparative experiments for software security. It is not enough to just collect
the data once, it is also important to have the possibility to assess the
results of the experiment. Therefore, \TestREx includes functionalities for
automatically collecting raw statistics on successes and failures of exploits.
We summarize the discussed tools and approaches in Table~\ref{tab:tools}.

\begin{table}[!ht]
\caption{Security testing and experimentation tools}
\label{tab:tools}
\longcaption{\columnwidth}{
	The existing tools and approaches provide various functionalities with respect
	to deployment (\eg, from running on a local virtual machine to providing
	controlled environments on real hardware). Most security research testbeds focus
	on the network level, while on the application level there are significantly
	less experimental frameworks.
\\
}
\begin{tabular}{| m{2.5cm} | m{5cm} | m{4.5cm} |}
\hline
	\textbf{Tool} &
	\textbf{Description} &
	\textbf{Exploit types} \\
\hline
	BugBox~\cite{nilson2013} &
	A corpus and exploit simulation environment for PHP web application
    vulnerabilities. &
	Selenium and Metasploit scripts in Python that exploit PHP application
    vulnerabilities. \\  	
\hline
	MalwareLab~\cite{allodi2013malwarelab} &
	A controlled environment for experimenting with malicious software. &
	Programs that exploit various software vulnerabilities or malware kits. \\
\hline
	MINESTRONE \cite{Evans2014} &
	A software vulnerability testing framework for C/C++ programs. The
    applications are deployed in virtualized environments via Linux Containers &
	Programs that exploit memory corruption, null pointer, number handling and
	resource leak vulnerabilities in C/C++ software. \\	  		
\hline
	DETER~\cite{benzel2011} &
	A testbed facility that consists of a large set (around 400) of real
    machines. The resources	infrastructure can be reconfigured on-the-fly upon
    request. &
	Programs that exploit various software vulnerabilities or malware kits. \\
\hline
	ViSe~\cite{arnes2006} &
	A virtual testbed for reproducing and collecting the evidence of security
    attacks that is based on VMWare virtualization environment. &
	Multi-level attacks that include network tampering and software
    vulnerability exploitation. \\
\hline
	SecuBat~\cite{kals2006} &
	Web vulnerability scanner that automatically scans live web sites for
    vulnerabilities using a web crawler infrastructure. &
	Specially crafted HTTP requests that exploit SQLi and XSS vulnerabilities. \\	
\hline
	vGrounds~\cite{jiang2006} &
	A virtual playground for malware assessment, that is created on top of a
    physical infrastructure - a machine, a cluster or a multi-domain overlay
    infrastructure. &
	Malicious software such as virtual worms or malware kits. \\
\hline
\end{tabular}
\end{table}

Nowadays, many exploits are publicly available in websites such as Packet
Storm\footnote{\url{https://packetstormsecurity.com/}} and
Exploit-DB\footnote{\url{https://www.exploit-db.com/}}, or even integrated in
penetration testing and exploitation frameworks such as
Metasploit\footnote{\url{https://www.metasploit.com/}} and
w3af\footnote{\url{http://w3af.org/}}. There is also a growing black market for
zero-day vulnerabilities and exploits~\cite{allodi2016} (these are
vulnerabilities for which no patch is available yet). These exploits are
usually provided ``as is'' and their reliability, i.e., their success rate
against different targets, can vary widely. Exploit reliability has been
studied in~\cite{dondo2015,holm2017} and the conclusion of both studies is that
most exploits have a low success rate when used ``off-the-shelf''. Both studies
were conducted for binary exploits (based on, e.g., buffer overflows). We are
unaware of any work studying the reliability of web application exploits.
Although we do not focus on increasing exploit reliability in this paper,
\TestREx can be used for testing it against web applications running with
different configurations in different software environments.

Continuous system testing and quality control are also related to \TestREx,
since they can be used to automate regression testing and testing of different
(evolving) versions of an application. Windmuller et al.~\cite{windmuller2013}
developed Active Continuous Quality Control (ACQC), an approach that uses
automata learning to infer the behavior of web applications. Neubauer et
al.~\cite{neubauer2014} extended ACQC to support risk-based 
testing~\cite{Felderer2014} by steering the ACQC process to increase risk
coverage. \TestREx differs from these works because it does not employ model 
inference, as the application and the test cases (exploits) are given by the 
user. The integration of evolving model inference techniques into our framework 
is an interesting venue of future research.

\section{Overview of \TestREx}
\label{sec:overview}

\TestREx was designed to provide testers with a convenient environment for
automated, large-scale experiments. We believe that \TestREx is useful for
developers as well. To support this claim, we give an example of a possible
loophole in a bug fixing workflow of a hypothetical company:
\begin{itemize}
	\item A tester finds a bug and opens a new issue in a bug tracking
    system. She submits it as a test case described in natural language,
    explaining all preconditions and steps needed to reproduce the bug.

	\item A manager assigns the issue to a developer. In order to pinpoint
    the source of the bug and understand how to fix it, the developer must
    reproduce the test case in his own setting. If the tester makes a mistake
    while creating the test case, the developer will be unable to trigger the
    bug. As a consequence, the developer rejects the fix request.

	\item In the worst case, it might take a long time before the bug will be
	re-discovered and eventually fixed. In a better case, more resources are
    wasted if the tester has to re-describe the bug, and a manager has to
    re-assign the bug to a developer.
\end{itemize}

Using \TestREx, the tester could create an ``executable description'' of a bug
in the form of a script, and a packed execution environment that allows to
instantly replay the situation that triggered the bug. Despite taking longer
for the tester to initially describe the bug this way, it has many advantages
over the natural language approach. First, the tester and the developer are
ensured that the bug can be reproduced. Second, the test case can be kept as a
template on which future tests can be developed, i.e., the first test is harder
to describe, but future tests can reuse parts of the first one. Third, the test
can be automatically added to a library of regression tests, to ensure that the
same bug will be detected if reinserted in future versions of the application.

\subsection{Terminology}
\label{sec:terminology}

Before we proceed, we introduce several general concepts that we use for
further discussion\footnote{Technically, these concepts are implemented using
Docker (\url{https://www.docker.io/}) -- we describe the implementation in
Section \ref{sec:implementation}. However, a different implementation may be
obtained using traditional virtual machines to which these general concepts can
be applied as well.}:

\begin{itemize}
	\item \texttt{Image} -- a snapshot of an application configured to run in
    a certain software environment, including this software environment. An
    image can be instantiated into a container that a tester can interact
    with.

	\item \texttt{Configuration} -- Configurations are used for creating
    images. We use this term as an intuitive meaning of a particular setup of
    an application and its supporting software components with particular
    values of setup parameters (configuration files, packages, etc.), as well
    as a set of instructions that are automatically executed in order to
    create an image in which these applications and components are
    ``deployed''.

	\item \texttt{Container} -- an instance of an image. This instance
    represents a certain state of an application and its software
    environment, that can be ``run'' for testing, and dismissed when testing
    is over. It can be either started using the pristine state of its base
    image (creating a new container, \ie, instance), or resumed from a
    certain existing state (re-using a container, that was already
    instantiated).
\end{itemize}

\subsection{Typical workflow}

An automated testbed should help security researchers in answering
(semi) automatically a number of security questions. Given an exploit $X$ that
successfully subverts an application $A$ running on an environment $E$:
\begin{compactenum}
	\item Will $X$ be successful on application $A$ running on a new
    environment $E'$?

	\item Will $X$ be successful on a new version of the application, $A'$,
    running	on the same environment?

	\item Will $X$ also be successful on a new version of the application,
    $A'$, running on a new environment $E'$?
\end{compactenum}
	
These questions can be exemplified in the following situation:
\begin{example}
\label{example:wordpress}
	We have a working SQL injection exploit for the WordPress 3.2 application
	running with MySQL and, we would like to know whether (1) the same exploit
    works for WordPress 3.2	running with PostgreSQL; (2) the same exploit works
    for WordPress 3.3 running	with MySQL; and (3) the same exploit works for
    WordPress 3.3 and PostgreSQL. \qed
\end{example}
We use this example throughout the paper to illustrate the concepts and
components used in the framework.

\begin{figure}[!ht]
\centering
\includegraphics[width=.75\textwidth,height=.7\textwidth]{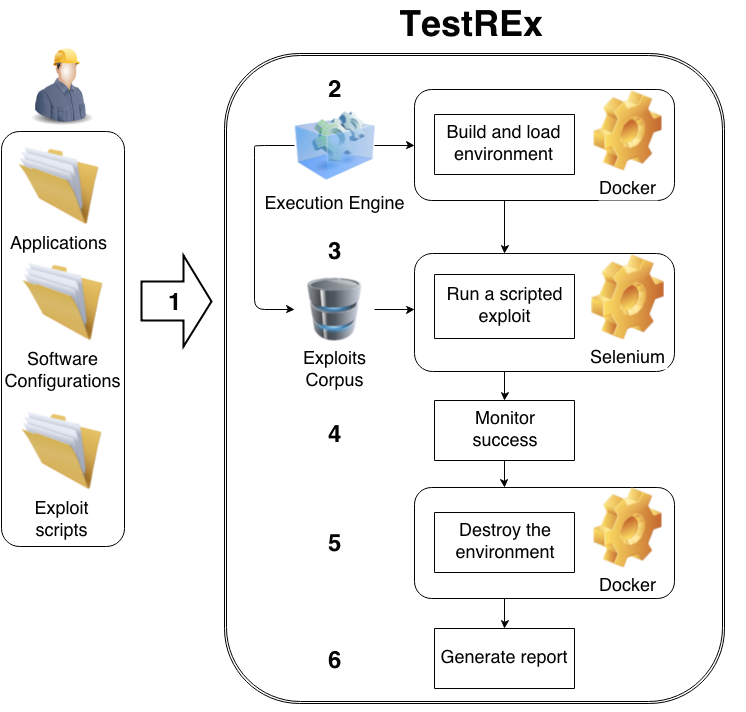}
\longcaption{\columnwidth}{
\\
	The workflow of \TestREx\ is straightforward: a tester provides
    configuration details of an application, its deployment environment, as
    well as the exploit scripts; \TestREx\ automates the remaining actions,
    such as building and loading the environment, and running and monitoring
    the exploit.
}
\caption{\TestREx workflow }
\label{fig:workflow}	
\end{figure}

A key feature that we have sought to implement, is that the architecture of
\TestREx should be easily extensible to allow for the inclusion of new
exploits, applications, and execution environments. Figure \ref{fig:workflow}
shows a typical workflow when an application and the corresponding scripted
exploits are deployed and run within \TestREx:
\begin{compactenum}	
	\item A tester provides the necessary \texttt{configuration} for a specific
	\texttt{image}, including the application and software component files, and
    the scripted exploits to be executed (the latter is optional, as \TestREx
    also supports manual testing).

	\item The \texttt{Execution Engine} component of \TestREx builds the
	\texttt{image} and instantiates the corresponding \texttt{container}.

	\item The \texttt{Execution Engine} runs corresponding exploit(s) against
    the	application container,

	\item and monitors whether the exploit execution was successful.

	\item \label{point:dismiss:container} After the exploit(s) are executed,
    the \texttt{Execution Engine} dismisses the corresponding container
    (optionally, further exploits may reuse the same container when the tester
    wishes to observe the cumulative effect of several exploits) and cleans up
    the environment.

	\item The exploit(s) execution report is generated.
\end{compactenum}

One of the main goals of \TestREx is to make the testing process as automated
as possible. Another important task is to make it possible to run applications
and exploits in a clean and isolated environment. Therefore, we included the
option of running every test against a clean state of the application. This
gives the possibility to run tests in parallel (see the point
\ref{point:dismiss:container} above).

\TestREx also includes some additional utilities. For instance, the
\texttt{Packing Module} allows to package configurations in compressed archive
files that can be easily deployed in another system running \TestREx. Also, the
\texttt{Utilities} module includes a collection of scripts to import
applications and exploits from other sources, such as BugBox, and to manage
images and containers.
	
\begin{example}
\label{example:components}
The inputs for Example~\ref{example:wordpress} are instantiated as
follows:
\begin{itemize}[noitemsep]
	\item \textbf{Application:} There are two applications of interest, each
    one is a set of \texttt{.html}, \texttt{.php} and \texttt{.js} files in a
	{\texttt{Wordpress}} folder.

	\item \textbf{Configuration:} There are four interesting configurations,
    one for	WP3.2 with MySQL, one for WP3.3 with MySQL, one for WP3.2 with
    PostgreSQL, and one for WP3.3 with PostgreSQL.

	\item \textbf{Image:} There are two possible images, one with Ubuntu
    Linux distribution, Apache web server and MySQL database engine, and one
    with Ubuntu, Apache and PostgreSQL.

	\item \textbf{Exploit(s):} There is only one exploit -- a script that
    navigates to the vulnerable web page, interacts with it and injects a
    payload, simulating the actions of an attacker. \qed
\end{itemize}
\end{example}

In our setting, exploits are unit tests: (1) every exploit is self-contained
and can be executed independently; and (2) every exploit is targeted to take
advantage of a specific vulnerability in a given application.

When using the framework in a specific application, the exploit can be written
by the tester or taken from a public source. In any case, the exploit code must
be compliant with what we expect from an exploit, \eg, it must be a subclass of
the \texttt{BasicExploit} class provided with \TestREx, and contain metadata
that specifies the target image and describes the exploit script (more details
are in Section~\ref{sec:creating:exploit} in the Appendix).	

Aegis~\cite{codaspy2017} extends the \TestREx architecture to test run-time
monitors, enforcing control-flow and data-flow integrity, as well as
authorization policies in workflow-driven web applications. The synthesized
monitors are deployed as Docker containers, and tests are implemented as
Selenium scripts -- as we have illustrated in Figure
\ref{fig:workflow}\footnote{The steps of Aegis are: (1) start new containers
with the appropriate applications; (2) run the workflows by using the Selenium
script; (3) repeat the workflows with monitoring on; (4) capture results. At
the end of the test session, the container can be destroyed and a new one
re-created if workflows' history are accounted for, and a pristine starting
image is important for repeatability.}.

\section{Implementation}
\label{sec:implementation}

\TestREx is implemented in Python, mainly because it allows fast and easy
prototyping and because of the availability of libraries and frameworks, such
as \texttt{docker-py} to interface it with Docker (see below). Below we
describe in details the implementation of each component of the framework.
	
\subsection{Execution Engine}
\label{sec:execution-engine}

The \texttt{Execution Engine} is the main \TestREx module that binds all its
features together. It supports three modes of operation: \emph{single},
\emph{batch} and \emph{manual}.

The \emph{single mode} allows testers to specify and run a desired exploit
against a container that corresponds to the chosen application image just once.
This is useful when the tester wants to quickly check whether the same exploit
works for a few different applications, different versions of the same
application or the same application deployed in different software
environments. A ``\texttt{.csv}'' report is generated at the end of the run.

To run applications and exploits in the \emph{batch mode}, \TestREx loops
through a folder containing exploit files, and runs them against respective
containers, generating a summary ``\texttt{.csv}'' report in the end. In this
mode, the \texttt{Execution Engine} maps exploits to application images by
scanning the metadata in each exploit, where appropriate target images are
specified by the tester.

For \emph{manual} testing, the \texttt{Execution Engine} instantiates a
container based on the chosen application image, and returns the control to the
tester (\eg, by opening a web browser and navigating to the application, or
returning a shell). No report is generated in this case.

The \texttt{Execution Engine} contains an additional setting for handing
containers when chosen exploits are executed: it is possible to either destroy
a particular container after the execution, in order to start with a ``fresh''
instance of the image for each exploit run; or to reuse the same container when
its state has to be preserved, so that further exploits may have a cumulative
effect that the tester wishes to observe.

\subsection{Applications}
\label{sec:applications}

Applications are packaged as ``\texttt{.zip}'' files containing all their
necessary code and other supporting files, such as database dumps. Unpacked
applications must be located under the ``{{\path{<testbed_root>/data/targets/
    applications}}}'' folder to be accessible by the \texttt{Execution Engine}.
	
As an example, we provide some applications with known vulnerabilities (they
are shown in Table \ref{tab:details:apps}, and their corresponding
vulnerability types are listed in Table \ref{tab:numbers:exploits}), most of
which are known real-world applications, only some of them being small
artificial examples developed by us to explore security vulnerabilities typical
for server-side JavaScript applications. 	

\subsection{Images and Containers}
\label{sec:containers}

Ideally, security testers should have the possibility of using various types of
computing components and platforms, regardless of the type of underlying
hardware and software that may be available.

To provide testers with the possibility of running applications in various
environments in a flexible, scalable, and cost-effective manner, we employ
software images (that are, implementation-wise, Docker images). Every such
image represents a data storage for virtualized computing components or
platforms, \eg, operating systems, application servers, database management
systems, and other types of supporting applications.

Instead of creating virtual machines for applications and their software
environments, we instantiate and run containers from corresponding images.
These containers are based on the
OCI\footnote{\url{https://www.opencontainers.org/}} standards, which are
nowadays widely accepted in industry as a form of ``lightweight
virtualization'' at the operating system level. They are sandboxed filesystems
that reuse the same operating system kernel, but have no access to the actual
operating system where they are deployed.

Some initial developments in this area were FreeBSD
Jails\footnote{\url{https://www.freebsd.org/doc/handbook/jails.html}}, Solaris
Zones\footnote{\url{https://docs.oracle.com/cd/E18440_01/doc.111/e18415/chapter_zones.htm}},
and Linux Containers\footnote{\url{https://linuxcontainers.org/}}. Currently,
Docker\footnote{\url{https://www.docker.io/}} is the \emph{de facto} standard
for containers. Docker provides a format for packing and running applications
within lightweight file repositories that are called Docker containers. We use
Docker to create images and instantiate containers.
	
Images are specified in Dockerfiles (a format defined by the Docker project) --
these files represent \emph{configurations} to which we refer in Section
\ref{sec:terminology}. Downloading generic software components and re-creating
a Docker container from a corresponding image every time an application has to
be run might be resource- and time-consuming. Therefore, we use image
inheritance supported for Dockerfiles, creating several images for containers
that hold generic software components, and can be reused by certain types of
web applications. For instance, such images may encapsulate an operating
system, a web server and a database engine, and their corresponding containers
are instantiated only once. We provide some predefined images for common
environments, using software components shown in
Table~\ref{tab:containers:example}. We use the following naming convention for
such images:
``{\path{<operating_system>-<webserver>-<database>-<others>}}''.
In contrast, for images which actually contain an application to be tested
(apart from generic software components) we use a different naming convention:
\\``{\path{<application-name>__[software-image-name]}}''.
	
When the \texttt{Execution Engine} invokes an application image, the
corresponding container will be instantiated and run using Docker. Then,
depending on run settings (see Section \ref{sec:execution-engine}), the
container will be handled correspondingly when chosen exploits are executed
(either destroyed, or reused for further exploit runs).

\begin{table}[!ht]
\centering
\caption{Software components for generic images currently provided with \TestREx}
\longcaption{\columnwidth}{\\}
\label{tab:containers:example}		
\begin{tabular}{| c | c | c |}
\hline
	\textbf{Web server} & \textbf{DB engine} & \textbf{OS} \\
\hline
	Apache & MySQL & Ubuntu \\
\hline
	Node.js & MySQL & Ubuntu \\
\hline
	Node.js & MongoDB & Ubuntu \\
\hline
	Tomcat & MySQL & Ubuntu \\
\hline
\end{tabular}
\end{table}

\begin{figure}[!ht]
\centering
\includegraphics[width=.72\columnwidth]{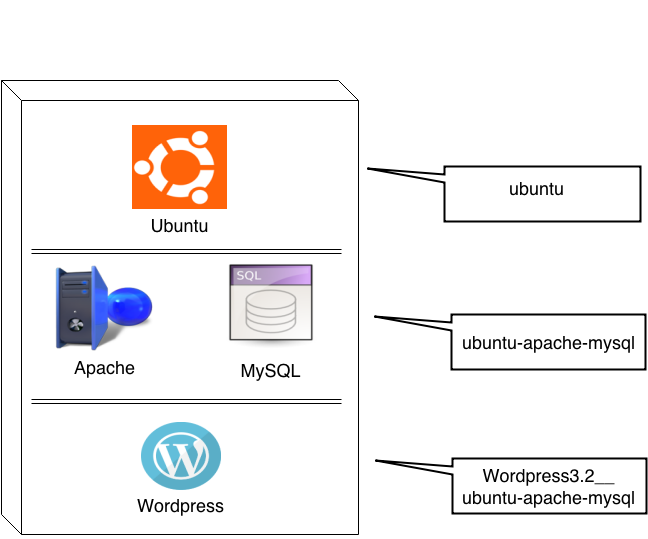}
\longcaption{\columnwidth}{
\\
	Application images are composed of several ``layers'': an operating system,	
    a web server, and a database engine -- the application itself is deployed
    on top. These components can be combined in all possible configurations
    supported by the application.
}
\caption{\emph{Wordpress3.2\_\_ubuntu-apache-mysql} image}
\label{fig:wp:container}	
\end{figure}

Figure~\ref{fig:wp:container} gives an intuition on how an image for the
\emph{WordPress 3.2} application can be composed with Dockerfiles: the image is
created on the basis of two images combining Ubuntu OS with Apache web server
and MySQL database.

\subsection{Configurations}
\label{sec:configurations}

Implementation-wise, configurations correspond to the contents of Dockerfiles
and supporting scripts that specify how an application can be installed and run
in a container, including, \eg, prerequisites such as preloading certain data
to a database, creating users, and starting a server. Additionally,
configuration data for applications may include databases and application data.

The configuration files must be placed in a separate folder under the
configurations root folder
(``{\path{<testbed_root>/data/targets/configurations}}''). We use the following
naming convention to simplify matching configuration files with images that can
be created using them:
``{\path{<app-name>__<app-container-name>}}''.
	
\begin{example}
A configuration folder for the application ``{\path{Wordpress_3.2}}'', might
have the names ``{\path{Wordpress_3.2__ubuntu-apache-mysql}}'' or
``{\path{Wordpress_3.2__ubuntu-apache-postgresql}}'', depending on the image
that is intended for it. \qed
\end{example}

Listings~\ref{listing:docker:example}
and~\ref{listing:shellscript:example} present an example of a Dockerfile and a
``\texttt{run.sh}'' file, used to configure a \emph{WordPress 3.2} application
within the ``\texttt{ubuntu-apache-mysql}'' image.

\begin{lstlisting}[frame=bt, basicstyle=\scriptsize, caption ={Dockerfile
example}, label={listing:docker:example}]
FROM ubuntu-apache-mysql	
RUN mkdir /var/www/wordpress
ADD . /var/www/wordpress	
RUN chmod +x /var/www/wordpress/run.sh
CMD cd /var/www/wordpress && ./run.sh
\end{lstlisting}	

In Listing~\ref{listing:docker:example}, line 1 specifies that the
image for this application is built on top of the
``\texttt{ubuntu-apache-mysql}'' image. In lines 2 and 3, the application files
are copied to the ``\texttt{/var/www/wordpress}'' folder in the image and in
lines 4 and 5, the ``\texttt{run.sh}'' script is invoked inside the container.
	
\begin{lstlisting}[frame=bt, basicstyle=\scriptsize, caption ={Shell script
file example}, label={listing:shellscript:example}]
#!/bin/bash
mysqld_safe &
sleep 5
mysql < database.sql
mysqladmin -u root password toor
apache2ctl start
\end{lstlisting}

In Listing~\ref{listing:shellscript:example}, lines 2-5 are used to start the
database server and pre-load the database with application data. Line 6 starts
the Apache web server.

\subsection{Exploits}
\label{sec:exploits}

Table 4 shows the classification of typical security flaws that might be present
in both client- and server-side parts of a web application\footnote{This
classification is according to the OWASP TOP 10:
\url{https://www.owasp.org/index.php/Top_10_2013-Top_10}}, which can be tested
with \TestREx.  The following security flaws may be present in web applications
regardless of their implementation and deployment details. Yet, their successful
exploitation strongly depends on the actual variant of deployment (\eg, MongoDB
versus MySQL database, type and the version of the web server, etc.).
\begin{small}
\begin{longtable}{|p{3cm}|p{7.5cm}|p{1.5cm}|}
\caption{Security flaws of web applications}\\
\hline
\centering \textbf{Security flaw}  & \centering \textbf{Description} &
	\parbox[t]{\columnwidth}{\textbf{Technical\\impact}} \\
\endhead
\hline
	\parbox[t]{\columnwidth}{SQL/NoSQL\\injection\\(SQLi/NoSQLi)}
	&
	User input is used to construct a database query and is not properly
    sanitized, allowing a malicious user to change the intended database query
    into an arbitrary one.
    \textbf{Threats: Information Disclosure, Data Integrity, Elevation of
    Privileges}.
	&
	Severe
	\\
\hline
	\parbox[t]{\columnwidth}{Code injection}
	&
	Similar to SQLi/NoSQLi, however, instead of a database, user input is
    executed by a code/command interpreter. Malicious payload can be executed
    on both client and server, and may result into a complete takeover of the
    host machine on which the vulnerable application runs.
	\textbf{Threats: Information Disclosure, Data Integrity, Elevation of
    Privileges, Host Takeover}.
	&
	Severe \\
\hline
	\parbox[t]{\columnwidth}{Cross-site scripting\\(XSS)}
	&
	Each time a user-supplied data is being displayed in a web browser, there
    is a risk of XSS attacks: attacker can supply JavaScript code that either
    gets executed in a victim's browser and stealing victim's credentials or
    making actions on her behalf. Almost any source of data can be an attack
    vector (\eg, direct user input, data coming from a database, etc.).
	\textbf{Threats: Information Disclosure, Elevation of Privileges}.
	&
	Moderate
	\\
\hline
	\parbox[t]{\columnwidth}{Cross-site\\request forgery\\(CSRF)}
	&
	CSRF attacks take advantage of benign applications that allow attackers to
    act on their behalf: user is secretly redirected from a trusted page to
    attacker's page, and user's authentication information is used by an
    attacker. Applications that allow manipulations with DOM container of its
    pages are vulnerable.
	\textbf{Threats: Session/Credentials Hijacking}.
	&
	Moderate
	\\
\hline
	\parbox[t]{\columnwidth}{Unvalidated\\URL redirects}
	&
	URL redirects instruct the web browser to navigate to a certain page. While
    this feature can be	useful in many different contexts, developers should be
    careful and restrict user manipulations with a destination page: an
    attacker may conduct phishing attacks using a trustworthy website that has
    this vulnerability.
	\textbf{Threats: Open Redirect}.
	&
	Moderate
	\\
\hline
	\parbox[t]{\columnwidth}{Sensitive data\\disclosure}
	&
	Sensitive/Personal data is attractive for attackers per definition,
    therefore the goal of most of attacks is to get a piece of such data. Since
    personal data is usually protected by law regulations, every such data flow
    in a web application must be protected against injection and interception
    attacks, as well as overly detailed error messages and application logic
    flaws that disclose context information to potential attackers.
	\textbf{Threats: Information Disclosure}.
	&
	Severe
	\\
\hline
	\parbox[t]{\columnwidth}{Test code leftovers}
	&
    A tester may insert a piece of testing code into the application and forget
    to remove it upon release. This can lead to any kind of unexpected
    behavior: for example, anyone could get access to the application with a
    login 'Bob' and a password `123' gaining full administrator access. Such
    forgotten pieces of test code are indistinguishable from maliciously
    crafted backdoors per se.
	\textbf{Threats: Backdoor}.
	&
	Severe
	\\
\hline
	\parbox[t]{\columnwidth}{Using known\\vulnerable\\components} &
	If vulnerable versions of third-party components are used (\eg, an
    open source library) in a web application, an attacker can identify known
    vulnerabilities and perform a successful attack. In many cases, developers
    are not aware of all components they are using for their application.
    Vulnerable component dependencies aggravate the problem.
	\textbf{Threats: Potentially all of the above}.
	&
	May vary
	\\
\hline
\end{longtable}
\label{tab:webapp:flaws}
\end{small}

In \TestREx, exploits may be any executable file that, when executed in a
specified context, provide testers with unauthorized access to, or use of
functionality or data within that context. Exploits include any sequence of
steps that must be taken in order to cause unintended behavior through taking
advantage of a vulnerability in an application and/or surrounding environment.
For example, exploits may be used to provide access to sensitive data, such as
financial data or personal data. Exploits may hijack capabilities or other
functionalities of applications and cause the applications to perform tasks
that are not desired by authorized users, such as tracking user activities and
reporting on these to the unauthorized user of the exploit. Other types of
exploits may allow unauthorized users to impersonate authorized users.

Still, the above description of an exploit is quite vague, and may lead to
having many automated exploit scripts that are not compatible due to various
differences in their implementation (\eg, as a consequence it may be difficult
to run them in a batch, and/or use them to produce a unified testing report).
To avoid these potential problems, we implemented exploits as a hierarchy of
Python classes that have the following minimal set of properties: (1) every
exploit contains metadata describing its characteristics such as name,
description, type, target application and container; (2) exploit classes must
pass logging information and results of the run to the \texttt{Execution
Engine}, providing a way for the \texttt{Execution Engine} to know that the
exploit execution was successful.

We also incorporate the \emph{Selenium Web
Driver}\footnote{\url{http://docs.seleniumhq.org/projects/webdriver/}} for
implementing exploits, as it can be used to simulate user/attacker actions in a
web browser, and provides all necessary means to automate them. Additionally,
it supports JavaScript execution and DOM interaction~\cite{nilson2013}. Every
Selenium-based exploit in the framework is a subclass of the
\texttt{BasicExploit} class, which encapsulates basic Selenium functionality to
automate the web browser (\eg, ``\texttt{setUp()}'' and ``\texttt{tearDown()}''
routines, logging and reporting, etc.). To create a new exploit, the tester has
to create a new exploit class, specify the exploit-specific metadata and
override the ``\texttt{runExploit()}'' method by adding a set of actions
required to perform an exploit. The success of an exploit run is also verified
within the ``\texttt{runExploit()}'' method - this might be different for every
exploit. This allows us to handle complex exploits that are not always
deterministic, such as heap spraying. For such cases, the exploit can be
specified to run a certain number of times until it is considered a success or
a failure.

\begin{table*}
\caption{Applications in the corpus}
\longcaption{\columnwidth}{
	The table shows the applications (real-world and artificial ones) that \TestREx
	currently includes.	The ``Base Images'' column specifies a base image which is
	used for creating a specific application image, and the ``Source'' column
	specifies the source from which we adapted exploits for these applications.
\\
}
\label{tab:details:apps}
\begin{tabular}{| m{1.6cm} | m{5.5cm} | m{3.2cm} | m{1.4cm} |}
\hline
	\textbf{Language} & \textbf{Applications} & \textbf{Base Images} & \textbf{Source} \\
\hline
	PHP &
	WordPress, CuteFlow, Horde, PHP Address Book, Drupal, Proplayer, Family
    Connections, AjaXplorer, Gigpress, Relevanssi, PhotoSmash, WP DS FAQ,
    SH Slideshow, yolink search, CMS Tree page view, TinyCMS, Store Locator
    Plus, phpAccounts, Schreikasten, eXtplorer, Glossword, Pretty Link &
	ubuntu-apache-mysql & 	
	BugBox \\

\hline
  	Java & WebGoat & ubuntu-tomcat-java &  WebGoat \\
\hline
	\parbox[t]{\columnwidth}{Server-side\\JavaScript} &
	CoreApp, JS-YAML, NoSQLInjection, ODataApp, SQLInjection, ST, WordPress3.2,
    XSSReflected, XSSStored &
	\parbox[t]{\columnwidth}{ubuntu-node,\\ubuntu-node-mongo,\\ubuntu-node-mysql} &
	\parbox[t]{\columnwidth}{Our\\examples} \\  	
\hline
\end{tabular}
\end{table*}

\subsection{Report}
\label{sec:report}

Different context conditions may transform failed exploit attempts into
successful ones, and vice versa. A given exploit test may include a number of
possible combinations of applications, execution environments, and exploits,
each of which may be configured in various ways. For example, an exploit that
may be successful in exploiting a first application in a first environment may
not be successful in exploiting that same application in a second environment,
but may be successful in exploiting a second application in the second
environment. Moreover, upon determining a success of a given exploit, it will
be necessary to make some change to the application and/or execution
environment, which will necessitate yet another testing (re-testing) of the
previously successful exploit to ensure that the change will prevent future
successful exploits.

Therefore, we include a reporting functionality: whenever \TestREx runs an
exploit, it generates a report that contains the information about its
execution. A report is a ``\texttt{.csv}'' file that the \texttt{Execution
Engine} creates or updates every time it runs an exploit. Every report contains
one line per exploit that was executed. This line consists of the exploit and
the target application names, identifier of an application-specific container,
the type of the exploit, the exploit start-up status, the exploit execution
result, and a comment field that may include other information that might be
exploit-specific. Along with this report, the \texttt{Execution Engine}
maintains a log file that contains information which can be used to debug
exploits.

Notice that this reporting functionality is also considered critical by bad guys
and is therefore present in almost all exploit kits available on black
markets~\cite{kotov2013anatomy}.

\begin{example}
The listing below shows a single entry from the \emph{Wordpress\_3\_2\_XSS}
exploit that was run against the \emph{WordPress 3.2} application. \qed
\end{example}

\begin{lstlisting}[belowskip=-0.8 \baselineskip,frame=bt, language = sh,
showstringspaces=false, breaklines=true, caption={An example of the report file
entry after the exploit run}, basicstyle=\scriptsize]
Wordpress_3_2_XSS, Wordpress3.2, ubuntu-apache
-mysql, XSS, CLEAN, SUCCESS, SUCCESS, 30.345,
Exploits for "XSS vulnerability in WordPress app"
\end{lstlisting}

\section{Evaluation}
\label{sec:evaluation}

As a starting point in the evaluation of \TestREx, we successfully integrated 
10 example exploits from WebGoat~\cite{webgoat}, as well as the corresponding
vulnerable web applications. We also developed exploits for 7 specially crafted
vulnerable applications, in order to demonstrate different types of exploits
for SQL injection, NoSQL injection, Stored and Reflected XSS, Path Traversal
and Code Injection vulnerabilities in applications that rely on server-side
JavaScript: the examples for the latter two vulnerability types take advantage
of vulnerabilities discovered in Node.js
modules~\cite{nodesec2013,nodesec2014}.

\begin{table}
\centering
\caption{Number of exploits in the corpus}
\longcaption{\columnwidth}{
	The table lists the number of exploits in the current corpus of \TestREx,
    broken down by a vulnerability type and a programming language of the
    vulnerable portion of the source code that makes the exploitation possible.
\\
}
\label{tab:numbers:exploits}
\begin{tabular}{|l|r|r|r|}
	\hline
	\textbf{Exploit} & \textbf{\#PHP} & \textbf{\#Java} & \textbf{\#Server JS} \\
	\hline
		XSS & 46 & 2 & 3 \\
	\hline
		SQLi & 17 & 2 & 1 \\
	\hline
		Code injection & 7 & - & 1 \\
	\hline
		Auth. flaws & 4 & 3 & - \\
	\hline
		Info. disclosure & 2 & - & - \\
	\hline
		Local file incl. & 2 & - & - \\
	\hline
		CSRF & 2 & - & - \\
	\hline
		DoS & 1 & - & - \\
	\hline
		DB backdoor & - & 1 & - \\
	\hline
		Param. tampering & - & 2 & - \\
	\hline
		Path traversal & - & - & 1 \\
	\hline
\end{tabular}
\end{table}

As \TestREx also supports the possibility of importing applications and
exploits from other similar testbeds, we imported all exploits and
corresponding applications from BugBox~\cite{nilson2013}. We used an automated
script that copies the applications and exploits into the corresponding folders
under \TestREx, and creates identical configuration files for every imported
application, using Apache as a web server and MySQL as a database server. We
were able to run most of the BugBox~\cite{nilson2013} native exploits and
collect statistics without modifying their source code. However, we had to
create a specific base image (as specified in Table \ref{tab:details:apps}), as
well as application images for BugBox applications.

Table~\ref{tab:numbers:exploits} summarizes the types of exploits that we
tested in various applications using \TestREx: it shows that \TestREx supports
a variety of typical web application security flaws. This successful case study 
was instrumental for SAP to decide to put forward a patent application of the 
technology behind \TestREx~\cite{sabetta2015multi}.

We also used \TestREx for an edition of the \emph{Laboratory on Offensive
Technologies} course taught at the University of Trento, Italy. The goal was to
teach MSc students enrolled into the Computer Science program about web
application vulnerabilities and exploits by: (i) exploring the vulnerability
corpus provided with \TestREx, and (ii) enlarging the corpus with new exploits
developed from scratch by them, or ported from other sources. In the Appendix,
we show an example of one exploit added to the corpus and detail the process of
adding it. Adding an exploit consists of three steps: \emph{deploying an
application}, \emph{creating configuration files and building containers}, and
\emph{creating and running an exploit}. We had the students focus on
understanding the last step, the actual exploits, in the first moment, so that
in a second moment they could write exploits for any application they wanted.

Although some students completed the assignments using \TestREx, other students
enrolled in the course preferred to develop system exploits (the two tracks
were offered as possibilities at the beginning of the course). We were unable
to draw relevant statistics about how the use of \TestREx impacted the learning
process, since that was the first edition of the course and there were only
around 20 students in total (which is not statistically relevant, if we assume
the usual 30 as the cut-off number). Nevertheless, we can affirm that \TestREx
helped in the preparation and teaching of the course as a source of ready
examples that could be easily run and analyzed in students' laptops (with
varying configurations of hardware and software).
In the future, we intend to analyze the use of \TestREx with a larger number of
students and possibly professional penetration testers, following an approach
inspired by~\cite{ceccato2017}.

\section{Potential Industrial Applications}
\label{sec:usage:model}

There are several uses of \TestREx that we are exploring in an industrial
setting, covering different phases of the Secure Development
Lifecycle~\cite{howard06} (SDL), and fulfilling the needs of different
stakeholders. Below we summarize the activities in different phases of the SDL
that can benefit from using \TestREx. Part of this work has also been an object
of a US Patent~\cite{sabetta2015multi}.

\subsection{Preparation and Training}

\textbf{Part of a training toolkit.} Security awareness campaigns, especially
secure coding training, are commonly conducted in large enterprises, also in
response to requirements from certification standards. From our own experience
with \TestREx, we believe that writing exploits may be an effective way to
acquire hands-on knowledge of how security vulnerabilities work in practice and
how to code defensively in order to prevent them. To quickly create a large
corpus of artificially vulnerable applications for training purposes, it is
possible to start from well-known applications and use vulnerability injection,
as done in~\cite{fonseca2014,pewny2016}. This way, we can easily create
multiple examples for each category of vulnerabilities, with different levels
of complexity for detection or exploitation.

\subsection{Design}

\textbf{Security testing of cloud-based applications}. One valuable use of
\TestREx is for cloud-based applications. In this scenario, a Cloud Service
Provider (CSP) provides the platform on which an Application Provider (AP) may
run their applications. CSPs allow the same application to be provided on
different platforms. However, such variations in context correspond to
potential difficulties in ensuring reliable and complete security testing,
because successful protection against an exploit in one context may prove
unsuccessful in another context. In this setting, \TestREx can provide highly
adaptable, flexible, efficient, and reliable testing for different
configurations, without requiring highly-specialized knowledge or abilities on
the part of the security tester. For example, the security tester may be an
employee of a CSP, which may wish to assure its customer APs that a secure
platform is in place. In turn, the security tester may be part of the AP, who
wishes to obtain independent testing of one or more platforms or platform
providers.

\subsection{Implementation and Verification}

\textbf{Automated validation and regression testing.} As part of the software
development lifecycle, \TestREx can be used to check the absence of known
vulnerabilities or to perform regression tests to verify that a previously
fixed vulnerability is not introduced again. To this end, a corpus of exploits
and configurations is stored in a corporate-wide repository and is used to
perform automated tests all along the development cycle. In large corporations,
the results of these tests are part of the evidence needed in order to pass
quality assurance gates. Currently, much of the process to produce such
evidence relies on manual work, which increases cost, errors and
unpredictability of the process. \TestREx can be used to accelerate and improve
the effectiveness and the predictability of quality assurance processes.

\textbf{Support for penetration testing.} An important problem arising in
penetration testing of large systems is the complexity of setting-up and
reproducing the conditions of the target system -- typically involving many
hosts and software components, each of which may need to be configured in a
specific way. A key strength of our framework is the ability to capture these
configurations as reusable scripts; this requires a non-negligible effort, but
the results can be reused across different penetration testing sessions. This
has the advantage of providing automation, reproducibility, and the ability to
proceed stepwise in the exploration of the effect of different configurations
and versions of the software elements on the presence (or absence) of
vulnerabilities in the system.

\subsection{Release and Response}

\textbf{``Executable documentation'' of vulnerability findings.} When a
vulnerability is found in a product, the ability to reproduce an attack is key
to investigate the root cause of the issue and to provide a timely solution. It
is current practice to use a combination of natural language and scripting to
describe the process and the configuration necessary to reproduce an attack.
The results of this practice may be erratic and complicate the security
response. \TestREx exploit scripts and configurations can be seen as
``executable descriptions'' of an attack. The production of exploits and
configurations could not just be the task of the security validation department,
but also of external security researchers, for which the company might set up a
bounty program requiring that vulnerabilities are reported in the form of
\TestREx scripts.

\textbf{Malware analysis.} Malicious third-party applications, also known as
malware, are applications intentionally designed to harm their victims, by,
e.g., stealing information or taking control of the victim's computer. Malware
in general, and especially web malware, are known to react differently to
different environments (usually to avoid detection)~\cite{chengyu2012,lu2014}.
Containers provide safe and repeatable environments for malware analysts to run
their experiments. One possible use of TestREx is as a highly configurable
sandboxing environment, where malware analysts can run potentially malicious
applications in different configurations of an application to study its
behavior. Another possible use is as a honeypot generator.
Honeypots~\cite{chen2011} are intentionally vulnerable applications deployed on
a network to capture and study attacks.

\textbf{Security testing of third-parties components.} According to the Black
Duck study~\cite{blackduck2016} more than 65\% of proprietary software vendors
integrate Free and Open Source Software (FOSS) components into their
applications.  Since enterprise software typically has long maintenance and
support lifecycles, older versions of FOSS components must be supported as
well. When a new vulnerability in a FOSS component is discovered, the vendor
has to verify whether if affects customers that are using different versions
of the software applications which may also contain different (older) versions
of the FOSS component. However, in this setting, traditional security testing
with static and dynamic analysis tools and code reviews may be
complicated~\cite{li2009development,baca2009static,Sahin-SMR242}. In this
scenario, \TestREx can be used to test whether the customers whose versions of
software applications are relying on older versions of a FOSS component are
affected by a newly disclosed vulnerability in a much newer version of that
component (which may not be the case, see~\cite{esej2015}).

\section{Conclusions}
\label{sec:conclusion}

In this paper, we presented \TestREx, a Framework for Repeatable Exploits that
combines a way of packing applications and execution environments, automatic
execution of scripted exploits, and automatic reporting, providing an
experimental setup for security testers and researchers. \TestREx provides
means for the evaluation of exploits, as the exploits are reproduced in a
number of different contexts, and facilitates understanding of effects of the
exploits in each context, as well as discovery of potential new
vulnerabilities.

We also provide a corpus of applications and exploits, either adapted from the
existing works, or developed by us -- we collected it to test the variety of
applications and exploits that can be handled by \TestREx.

\subsection{Lessons learned}

We can summarize the key lessons learned during the design and development of
\TestREx as follows: (1) build on top of existing approaches; (2) have a simple
and modular architecture; (3) find reliable information on applications,
exploits and execution environments in order to replicate them.

Building on top of the existing work, like we did with BugBox~\cite{nilson2013}
for the format of our exploits, and MalwareLab~\cite{allodi2013malwarelab} for
the vulnerability experimentation design, was extremely valuable. This
simplified our design and development time, and allowed us to quickly add a
large corpus of applications and exploits on which we could test our
implementation.

The functionality that TestREx offers can be achieved to a certain
extent by separately using its individual tool components. For
example, a tester can use a regular virtual or physical machine (or plain
Docker) to deploy the software components of interest, and then either run a
Selenium script or perform manual testing. However, she will have to become
acquainted with all these tools and perform the experiments manually.

This approach only works if (i) the tester has only \emph{one single
configuration} to deploy and (ii) s/he has  \emph{one single vulnerability} to
test, and (iii) the tester is \emph{both a security expert and a functional
expert} (i.e.\ knows the application and the actually deployed configurations).
In all other cases, repeatedly using the individual tools would not scale.

Indeed, to create and test an exploit, a tester must first understand the
``mechanics'' of a vulnerability that can be exploited, or adapt existing
exploits to specific conditions. Also, in many cases, publicly available 
exploit descriptions are vague, limited to a proof-of-concept (which may not
necessarily work), and often lack information on how to reproduce them in a
specific software environment.

Then this must be adapted to the actual configurations that are used in the
company. This may be quite difficult due to the fact that the information on how
to configure a certain application environment may not be detailed enough in the
official documentation. Therefore, a functional expert may be more appropriate,
but s/he may not have the security skills needed to deploy the vulnerability.

Instead, TestREx provides a completely automated solution in which the 
knowledge of different aspects can be ``hidden away'' from its different users 
(e.g., security testers might only need to know how to write and run exploit 
scripts, and everything else will be just ``hooked up'' and executed right out 
of the box; a functional tester would only need to write the configuration and 
just use the exploit scripts as provided).

Large scale usage of TestREx requires an initial investment which lies in
creating a set of software configurations (images) that can be then reused
company-wide: as we learned during the evaluation of TestREx, creating reliable
software configurations is the most difficult part, while creating exploit
scripts is comparatively easy. However, once this initial effort is invested, 
it becomes extremely easy to add new exploit scripts, run them in different
combinations, combine different software images (we use Docker inheritance),
and collect reports.

\subsection{Future work}
We intend to extend the architecture of \TestREx to add support for plugins.
Plugins (e.g., proxy tools, vulnerability scanners) could be used to facilitate
activities such as pentesting, vulnerability analysis, and malware analysis,
mentioned in section~\ref{sec:usage:model}.

We would also like to expand the current vulnerability corpus by taking public
exploits from, e.g., Exploit-DB and reconstructing the vulnerable environments
in \TestREx.

\bibliographystyle{plain}
\bibliography{long-names,references}

\appendix

\section{Contributing to \TestREx}
\label{sec:contributing}

Here we describe in more detail the steps needed to add an experiment to
\TestREx, given an existing application. These steps consist of: adding an
application; creating configuration files for images; instantiating containers;
creating and running exploits. Again, we use \emph{WordPress 3.2} as the
example application.

\subsection{Deploying an Application}

The code of the application must be copied into a separate folder under the
applications root ``{\path{<testbed_root>/data/targets/applications}}''. The
folder name must correspond to a chosen name of the application in the testbed.

To deploy the \emph{WordPress 3.2} application, copy all of its files to the
folder ``{\path{<testbed_root>/data/targets/applications/WordPress_3_2}}''.
	
\subsection{Creating Configuration Files and Building Containers}

If there are no generic images that might be reused for creating a new image
for the application set up, this image must be created in the first place.
Configuration files for generic images are located under the
``{\path{<testbed_root>/data/targets/containers}}'' folder.

In our example, we create a generic image with the
\texttt{ubuntu-apache-mysql} name, since the application requires \emph{Apache}
as a web server and \emph{MySQL} as a database engine. To do this, we create a
Dockerfile under
``{\path{<testbed_root>/data/targets/containers/ubuntu-apache-mysql}}'' that
contains the code shown in Listing~\ref{listing:software:container}, and build
it with the script located under
``{\path{<testbed_root>/util/build-images.py}}''.

\begin{lstlisting}[frame=bt, breaklines=true, basicstyle=\scriptsize, caption ={
The Dockerfile for creating the \texttt{ubuntu-apache-mysql} generic image},
label={listing:software:container}]
FROM ubuntu:raring
RUN apt-get update
RUN DEBIAN_FRONTEND=noninteractive apt-get -y install mysql-client mysql-server apache2 libapache2-mod-php5 php5-mysql php5-ldap
RUN chown -R www-data:www-data /var/www/
EXPOSE 80 3306
CMD ["mysqld"]
\end{lstlisting}	

As a next step, we create configuration files for the image that will hold the
application, extending the above generic image. We create a new Dockerfile and
a shell script file under the
``{\path{<testbed_root>/data/targets/configurations/Wordpress_3_2__ubuntu-apache-mysql}}''
folder (see Listings~\ref{listing:docker:example}
and~\ref{listing:shellscript:example} in the Section~\ref{sec:configurations}
for the code examples).
	
There is no need to manually invoke Docker for instantiating a container based
on this image for running exploits or manual testing, as \texttt{Execution
Engine} does it automatically.

\subsection{Creating and Running an Exploit}
\label{sec:creating:exploit}

Finally, we create an exploit for the \emph{Wordpress 3.2} application by
creating a Python class under the ``{\path{<testbed_root>/data/exploits}}''
folder. As mentioned in the previous sections, to ensure integration with the
\texttt{Execution Engine}, the new exploit class must be a subclass of the
already existing \texttt{BasicExploit} class. As a last step, we specify the
exploit's metadata using the \texttt{attributes} dictionary, and specify the
steps required to run the exploit within the ``\texttt{runExploit()}'' method
(see Listing \ref{listing:exploit:example}).
	
\begin{lstlisting}[frame=bt, language=Python, basicstyle=\scriptsize, breaklines=true, showstringspaces=false, caption ={Wordpress\_3\_2\_Exploit.py file contents}, label={listing:exploit:example}]
from BasicExploit import BasicExploit
class Exploit(BasicExploit):
  attributes = {
      'Name' :        'Wordpress_3_2_XSS',
      'Description' : "XSS attack in Wordpress 3.2",
      'Target' :      "Wordpress3.2",
      'Container': 'ubuntu-apache-mysql',
      'Type' : 'XSS'
   }

   def runExploit(self):
      w = self.wrapper
      w.navigate("http://localhost:49160/wordpress/wp-admin/post-new.php?post_type=page")
      ```
      ```
      content_elt = w.find("content").clear()
      content_elt.keys("<script>alert(\"XSS!!\")</script>")
      w.find("publish").click()

      w.navigate("http://localhost:49160/wordpress/?page_id=23")
      alert_text = w.catch_alert()
      self.assertIn("XSS", alert_text, "XSS")
\end{lstlisting}

Listing~\ref{listing:exploit:example} shows the stored XSS exploit for the
\emph{Wodpress 3.2} application. The script navigates to the login page of the
\emph{Wordpress} application, logs in as the administrator (the full list of
steps is shortened in the listing for the sake of brevity), and creates a new
post putting the \path{<script>alert(`XSS`)</script>} string as the content. To
verify whether the exploitation was successful, the script navigates to the
newly created post and checks if an alert box with the ``XSS'' message is
present.

In order to test whether the same exploit would work for different older or
newer versions of the Wordpress application, a tester may reuse the
configuration files of Wordpress 3.2. She only needs to provide the application
files of that version. We indeed tested the exploit from the example on other
versions of Wordpress: 4.2.15 and 4.8 (the latest available version). It works
against the first, but does not the work against the latter, as this
vulnerability was fixed for the latest version.

Listing~\ref{listing:commands} shows the list of commands for
different running modes in \TestREx:
\begin{compactenum}
	\item To run the application container for \emph{manual} testing, a tester
    has to use the ``\emph{--manual}'' flag and the corresponding
    \emph{application} image. \TestREx will run the container and halt, waiting
    for the interrupt signal from the tester. In this mode, when the container
    is up, the application can be accessed from a web browser by navigating to
    ``http://localhost:49160''.

	\item In the \emph{single mode} a tester can select a specific exploit and
    run it against a specific \emph{application} image.

	\item In the \emph{batch mode for a single application}, a tester has to
    specify	the running mode as ``\emph{--batch}'', and select the desired
    \emph{application} image. \TestREx will invoke a Docker container for the
    image, search for the exploits that are assigned to the application
    (through exploits' metadata), and run all of them one by one.

	\item Finally, if a tester specifies nothing but the ``\emph{--batch}''
	running mode, \TestREx will invoke containers for all \emph{application}
    images that are currently in the corpus, and run all corresponding exploits
    against them.
\end{compactenum}

\begin{lstlisting}[frame=bt, language = sh, showstringspaces=false,
breaklines=true, basicstyle=\scriptsize, caption={Running modes in \TestREx},
label={listing:commands}]
#1: Manual mode
sudo python run.py --manual --image
		[app-name]__[image-name]

#2: Single exploit mode
sudo python run.py --exploit [exploit-name].py
		--image [app-name]__[image-name]

#3: Batch mode for a single application
sudo python run.py --batch
		--image [app-name]__[image-name]

#4: Batch mode for all applications
sudo python run.py --batch
\end{lstlisting}

By default, the exploit execution report is saved into the
``{\path{<testbed_root>/reports/ExploitResults.csv}}'' file. In order to
specify a different location for the results, the tester may add an additional
parameter to the \texttt{run} command:
\texttt{--results new/location/path.csv}.

\end{document}